\documentclass{article}
\usepackage{amsmath,euscript,amssymb,amsfonts,graphicx,bm}
  \usepackage{graphics}

\usepackage{color}

\newcommand{\R}{{\mathbb R}}

\newcommand{\calU}{{\mathcal U}}

\newcommand{\e}{\mathrm{e}}

\newcommand{\x}{\mathbf{x}}
\newcommand{\y}{\mathbf{y}}

\newcommand{\X}{\mathbf{X}}
\renewcommand{\P}{\mathbb{P}}

\newcommand{\n}{\mathbf n}

\newcommand{\E}{{\mathbb E}}

\begin{document}

\title{\bf Target competition for resources under diffusive search-and-capture} 
\author{ \em
P. C. Bressloff, \\ Department of Mathematics, 
University of Utah \\155 South 1400 East, Salt Lake City, UT 84112}

 \maketitle

\begin{abstract}
In this paper we use asymptotic analysis to determine the steady-state mean number of resources in each of $N$ small interior targets within a three-dimensional bounded domain. The accumulation of resources is based on multiple rounds of search-and-capture events; whenever a searcher finds a target it delivers a resource packet to the target, after which it escapes and returns to its initial position (resetting after capture). The searcher is then resupplied with cargo and a new search process is initiated after a random delay. Assuming that the accumulation of resources is counterbalanced by degradation, one can derive general expressions for the moments of the resource distribution. We use this to show that the mean number of resources in a target is proportional to its effective ``shape capacitance.'' We then extend the analysis to the case of diffusive search with stochastic resetting before capture, where the position of the searcher is reset to its initial position at a random sequence of times that is statistically independent of the ongoing search process, in contrast to the sequence of resetting times after capture. \end{abstract}

\
\maketitle

\section{Introduction}

Random search strategies are found throughout the natural world as a way of efficiently searching for one or more targets of unknown location. Examples include animals foraging for food or shelter 
\cite{Bell91,Bartumeus09,Viswanathan11}, proteins searching for particular sites on DNA \cite{Berg81,Halford04,Coppey04,Lange15}, biochemical reaction kinetics \cite{Loverdo08,Benichou10}, motor-driven intracellular transport of vesicles \cite{Bressloff13,Maeder14,Bressloff15}, and cytoneme-based morphogen transport \cite{Kornberg14,Stanganello16,Zhang19,Bressloff19}. Most theoretical studies of these search processes take a searcher-centric viewpoint, focusing on the first passage time (FTP) problem to find a target. In this paper we take a target-centric viewpoint, whereby one keeps track of the accumulation of resources in the targets due to multiple rounds of search-and-capture events together with degradation.
As we have previously shown within the specific context of cytoneme-based morphogenesis \cite{Bressloff19,Bressloff20a}, the steady-state distribution of resources accumulated by a set of targets can be determined by reformulating a search-and-capture model as a G/M/$\infty$ queuing process \cite{Takacs62,Liu90}. Queuing theory concerns the mathematical analysis of waiting lines formed by customers randomly arriving at some service station, and staying in the system until they receive service from a group of servers. A sequence of search-and-capture events can be mapped onto a queuing process as follows: individual resource packets are analogous to customers, the delivery of a packet corresponds to a customer arriving at the service station, and the degradation of a resource packet is the analog of a customer exiting the system after being serviced. Assuming that the packets are degraded independently of each other, the effective number of servers in the corresponding queuing model is infinite, that is, the presence of other customers does not affect the service time of an individual customer. One of the advantages of formulating the search problem as a queuing process is that one can use renewal theory to calculate moments of the distribution of resources in steady state. 

The structure of the paper is as follows. In section 2, we formulate the general problem, and give expressions for the steady-state mean and variance of the distribution of resources across a set of targets labeled $k=1,\ldots,N$. These depend on the splitting probabilities $\pi_k$ and conditional MFPTs $T_k$ associated with a single search-and-capture event. Here $\pi_k$ is the probability that the particle first finds the $k$-th target. Since, this probability is less than unity due to target competition, it follows that the MFPT to find the $k$-th target is infinite unless it is conditioned on successfully finding the given target, which yields the conditional MFPT $T_k$. We also allow for delays between successive search-and-capture events due to the time needed for a particle to load/unload resources. We then develop the theory by considering diffusive search in a three-dimensional (3D) bounded domain containing $N$ small interior targets (section 3). In particular, we use asymptotic analysis to show that the mean number of resources in a target is proportional to its effective ``shape capacitance,'' see \cite{Redner} for a definition. Finally, in section 4 we extend our analysis to diffusive search processes that also includes stochastic resetting before capture. The latter resetting protocol has attracted considerable attention within the statistical physics community, see the recent review \cite{Evans20} and references therein. The underlying idea is that the position of a particle performing a stochastic search for some target is reset to a fixed location at a random sequence of times, which is typically (but not necessarily) generated by a Poisson process. This sequence is statistically independent of the ongoing search process, in contrast to the sequence of resetting times following target capture. 

\section{Steady-state distribution of resources and $G/M/\infty$ queues}

\begin{figure}[b!]
\raggedleft
\includegraphics[width=10cm]{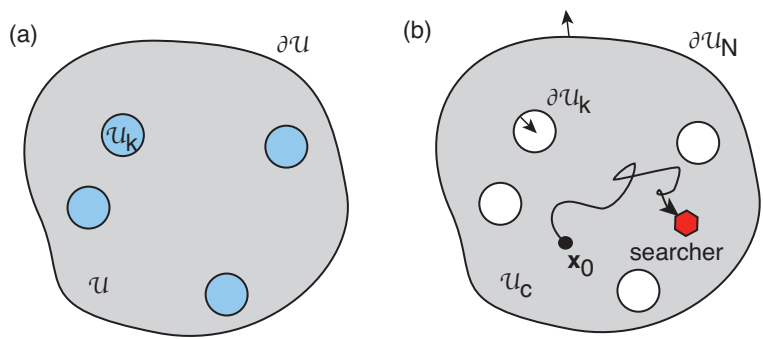} 
\caption{(a) Simply-connected bounded domain ${\mathcal U}$ containing $k=1,\ldots,N-1$ interior targets $\calU_k$ and an absorbing exterior boundary $\partial {\mathcal U}$. (b) Particle searching in the domain $\calU_c={\mathcal U}\backslash  \bigcup_{k=1}^{N-1}\calU_k$ with an absorbing boundary $\partial {\mathcal U}_c= \bigcup_{j=1}^{N}\partial{\mathcal U}_j$, $\partial{\mathcal U}_N=\partial {\mathcal U}$.}
\label{fig1}
\end{figure}

Consider a simply-connected, bounded domain ${\mathcal U}\subset \R^d$ with a set of $k=1,\ldots,N$ interior absorbing targets or traps $\calU_k\subset {\mathcal U}$, see Fig. \ref{fig1}(a). Furthermore, suppose that the boundary of the domain $\partial \calU$ is a totally absorbing exterior target. (For simplicity, we do not consider the more general case of mixed exterior boundary conditions, where only a subset of the boundary $\partial {\mathcal U}$ is absorbing and the complementary set is reflecting. If the boundary is totally reflecting then we only have interior targets, see section 3.) Within the context of cell biology $\calU$ could be identified with the cell cytoplasm, $\partial \calU$ with the cell membrane, and the interior targets with subcellular targets such as the cell nucleus or the endoplasmic reticulum. Introducing the multiply-connected domain $\calU_c={\mathcal U}\backslash \bigcup_{k=1}^{N-1}\calU_k$, it follows that the boundary of $\calU_c$ can be partitioned into a set of $N-1$ interior absorbing boundaries $\partial U_k$, $k=1,\ldots,N-1$, and a single exterior absorbing boundary $\partial U_N=\partial U$, that is, $\partial {\mathcal U}_c=\bigcup_{j=1}^N\partial{\mathcal U}_j$. Now suppose that a particle (searcher) is subject to Brownian motion in $\calU_c$ with $N$ totally absorbing targets corresponding to the $N$ components of the boundary $\partial \calU_c$.

The probability density $p(\x,t|\x_0)$ for the particle to be at position $\x$ at time $t$, having started at $\x_0$, evolves according to the diffusion equation 
\begin{equation}
 \label{1J}
\frac{\partial p(\x,t|\x_0)}{\partial t}=D\nabla^2p(\x,t|\x_0)=-\nabla\cdot {\mathbf J}(\x,t|\x_0),
\end{equation}
where ${\mathbf J}=-D\nabla p$ is the probability flux. This is supplemented by the boundary condition
\begin{equation}
 p(\x,t|\x_0)=0,\ \x \in  \bigcup_{j=1}^N\partial{\mathcal U}_j,
\end{equation}
and the initial condition $p(\x,0|\x_0)=\delta(\x-\x_0)$. 

Let ${\mathcal T}_k(\x_0)$ denote the FPT that the particle is captured by the $k$-th target, with ${\mathcal T}_k(\x_0)=\infty$ indicating that it is not captured.
Define $\Pi_k(\x_0,t)$ to be the probability that the particle is captured by the $k$-th target after time $t$, given that it started at $\x_0$:
\begin{equation}
\label{Pi}
\Pi_k(\x_0,t)=\P[t<{\mathcal T}_k(\x_0)<\infty ]=\int_t^{\infty} J_k(\x_0,t')dt',
\end{equation}
where
\begin{equation}
\label{Jk}
J_k(\x_0,t)=\int_{\partial \calU_k} {\mathbf J} (\sigma,t|\x_0)\cdot {\bf n} d\sigma.
\end{equation}
Note that the normal $\n$ to the boundary $\partial U_k$ is always taken to point from inside to outside the search domain, see Fig. \ref{fig1}. Moreover, differentiating equation (\ref{Pi}) and taking Laplace transforms implies that
\begin{equation}
\label{PiLT}
s\widetilde{\Pi}_k(\x_0,s)-\pi_k(\x_0)=-\widetilde{J}_k(\x_0,s).
\end{equation}
 The splitting probability $\pi_k(\x_0)$ and conditional MFPT $T_k(\x_0)$ for the particle to be captured by the $k$-th target are then
\begin{equation}
\label{pi}
\pi_k(\x_0)  =\Pi_k(\x_0,0)= \int_0^\infty J_k(\x_0,t) dt=\widetilde{J}_k(\x_0,s),
\end{equation}
and
\begin{equation}
T_k(\x_0) = \mathbb{E}[{\mathcal T}_k | {\mathcal T}_k < \infty]=\frac{1}{\pi_k(\x_0)}\int_0^\infty \Pi_k(\x_0,t) dt .
\end{equation}
We will assume that $\sum_k\pi_k(\x_0)=1$, which implies that the particle is eventually captured by a target with probability one. Finally, note that integrating equation (\ref{1J}) with respect to $\x$ and $t$ implies that the survival probability up to time $t$ is
\begin{eqnarray}
\label{q00}
Q(\x_0,t)&= \int_{\calU}p(\x,t|\x_0)d\x=\sum_{k=1}^N\Pi_k(\x_0,t).
\end{eqnarray}

Now suppose that, rather than being permanently absorbed or captured by a target on the boundary, the particle delivers a discrete packet of some resource to the target and then returns to $\x_0$, initiating another round of search-and-capture. We will refer to the delivery of a single packet as a capture event, and the return to $\x_0$ as resetting after capture. The sequence of events resulting from multiple rounds of search-and-capture leads to an accumulation of packets within the targets, which we assume is counteracted by degradation at some rate $\gamma$. We will assume that the total time for the particle to unload its cargo, return to $\x_0$ and start a new search process is given by the random variable $\widehat{\tau}$, which for simplicity is taken to be independent of the location of the targets. (This is reasonable if the sum of the mean loading and unloading times is much larger than a typical return time.) Let $n\geq 1$ label the $n$-th capture event and denote the target that receives the $n$-th packet by $j_n$. If ${\mathbb T}_n$ is the time of the $n$-th capture event, then the inter-arrival times are 
\begin{equation}
\Delta_n:={\mathbb T}_n-{\mathbb T}_{n-1}=\widehat{\tau}_n+{\mathcal T}_{j_n},\quad n\geq 1,
\end{equation}
with $\E[{\mathcal T}_{j}]=\pi_jT_j$.
Finally, given an inter-arrival time $\Delta$, we denote the identity of the target that captures the particle by ${\mathcal K}(\Delta)$. We can then write for each target $j$,
\begin{eqnarray}
\label{F1}
F_{j}(t)&= \P[\Delta <t,{\mathcal K}(\Delta)=j]= \P[\Delta <t,|{\mathcal K}(\Delta)=j]\P[{\mathcal K}(\Delta)=j] \\
&=\pi_{j} \int_0^{t}{\mathcal F}_{j}(\Delta)d\Delta ,\nonumber
\end{eqnarray}
where ${\mathcal F}_{j}(\Delta)$ is the conditional inter-arrival time density for the $j$-th target. Let $\rho(\widehat{\tau})$ denote the waiting time density of the delays $\widehat{\tau}_n$. Then
\begin{eqnarray*}
 {\mathcal F}_{j}(\Delta)&=\int_0^{\Delta}dt\int_0^{\Delta} d\widehat{\tau} \delta(\Delta -t-\widehat{\tau})f_{j}(t)\rho(\widehat{\tau})=\int_0^{\Delta} f_{j}(t) \rho(\Delta-t)dt,
\end{eqnarray*}
where $f_{j}(t)=J_j(t)/\pi_j$ is the conditional first passage time density for a single search-and-capture event that delivers a packet to the $j$-th target. (For notational simplicity, we drop the explicit dependence on the initial position $\x_0$.) In particular,
\begin{equation}
\label{cMFPT}
T_{j}=\int_0^{\infty} tf_{j}(t)dt =-\frac{1}{\pi_j}\int_0^{\infty} t\frac{d\Pi_{j}(t)}{dt}dt =\frac{\widetilde{\Pi}_j(0)}{\pi_j}.
\end{equation}
Laplace transforming the convolution equation then yields
\begin{equation}
\label{calF}
\widetilde{{\mathcal F}}_{j}(s)=\widetilde{f}_{j}(s)\widetilde{\rho}(s).
\end{equation}

As we have previously shown within the specific context of cytoneme-based morphogenesis \cite{Bressloff19,Bressloff20a}, the steady-state distribution of resources accumulated by the targets can be determined by reformulating the model as a G/M/$\infty$ queuing process \cite{Takacs62,Liu90}. Since the analysis carries over to diffusive search processes, we simply state the results for the steady-state mean and variance. Let $M_k$ be the steady-state number of resource packets in the $k$-th target. The mean is then
\begin{equation} \label{mean}
\overline{M}_k  =  \frac{\pi_k}{\gamma \sum_{j=1}^N \pi_{j} (T_j +\tau_{\rm cap} )}=\frac{\pi_k}{\gamma(T+\tau_{\rm cap})},
\end{equation}
where $\tau_{\rm cap}=\int_0^{\infty}\tau\rho(\tau)d\tau$ is the mean loading/unloading time and $T=\sum_{j=1}^N\pi_jT_j$ is the unconditional MFPT.
Equation (\ref{mean}) is consistent with the observation that $T +\tau_{\rm cap} $ is the mean time for one successful delivery 
of a packet to any one of the targets and initiation of a new round of search-and-capture. Hence, its inverse is the mean rate of capture events and $\pi_k$ is the fraction that are delivered to the $k$-th target (over many trials). (Note that equation (\ref{mean}) is known as Little's law in the queuing theory literature \cite{Little61} and applies more generally.) The dependence of the mean $\overline{M}_k$ on the target label $k$ specifies the steady-state allocation of resources across the set of targets. It will depend on the details of the particular search process (\ref{1J}), the geometry of the domain $\calU$, the initial position $\x_0$, and the rate of degradation $\gamma$. Similarly, it can be shown that
the variance of the number of resource packets is
\begin{eqnarray}
\label{var0}
 \mbox{Var}[M_k]=\overline{M}_k\left [ \frac{\pi_{k}\widetilde{\mathcal F}_{k}(\gamma)}{1-\sum_{j=1}^N\pi_{j} \widetilde{\mathcal F}_{j}(\gamma)}+1-\overline{M}_k
\right ]. 
\end{eqnarray}
Finally, noting that $\pi_{k}\widetilde{\mathcal F}_{k}(\gamma)=\widetilde{\rho}(\gamma)\widetilde{J}_{k}(\gamma)$ and using equations (\ref{q00}) and (\ref{Pi}) yields
\begin{eqnarray}
\label{var}
 \mbox{Var}[M_k]=\overline{M}_k\left [ \frac{\widetilde{\rho}(\gamma)\widetilde{J}_{k}(\gamma) }{1-\widetilde{\rho}(\gamma)[1-\gamma \widetilde{Q}(\gamma)]}+1-\overline{M}_k
\right ]. 
\end{eqnarray}
Although the mean $\overline{M}_k$ only depends on the quantities $\pi_k$ and $T_k$, the variance and higher-order moments involve the Laplace transformed fluxes $\widetilde{J}_{j}(\gamma)$, which are often more difficult to calculate.

Both the mean and variance vanish in the fast degradation limit $\gamma \rightarrow \infty$, since resources delivered to the targets are immediately degraded so that there is no accumulation. On the other hand, in the limit of slow degradation ($\gamma \rightarrow 0$), the mean and variance both become infinite. (There is no stationary state when $\gamma=0$.) Rather than working with the variance, however, it is more convenient to consider the Fano factor
\begin{equation}
\label{FF}
 FF_{k}=\frac{\mbox{Var}[M_{k}]}{\overline{M}_{k}}= 1+ \frac{\widetilde{\rho}(\gamma)\widetilde{J}_{k}(\gamma) }{1-\widetilde{\rho}(\gamma)[1-\gamma \widetilde{Q}(\gamma)]}-\overline{M}_k. 
 \end{equation}
It immediately follows that
\begin{equation}
\label{cong}
 \lim_{\gamma \rightarrow \infty} FF_{k}=1.
\end{equation}
In order to determine $FF_k$ in the limit $\gamma \rightarrow 0$, we consider the Taylor expansion of equation (\ref{calF}):
\begin{eqnarray*}
 \pi_j \widetilde{{\mathcal F}}_{j}(\gamma)&=& {\widetilde{J}_{j}(\gamma)} \widetilde{\rho}(\gamma)\\
 &=&\pi_j-\gamma \pi_j (T_j+\tau_{\rm cap})+\frac{\gamma^2\pi_j}{2}(T_j^{(2)}+\tau_{\rm cap}^{(2)}+2T_j\tau_{\rm cap}) +O(\gamma^3),
\end{eqnarray*}
and
\begin{eqnarray*}
\sum_{j=1}^N\pi_j \widetilde{{\mathcal F}}_{j}(\gamma)&=1-\gamma (T+\tau_{\rm cap}) +\frac{\gamma^2}{2}(T^{(2)}+2T\tau_{\rm cap}+\tau_{\rm cap}^{(2)})+O(\gamma^3),\nonumber 
\end{eqnarray*}
where $\tau_{\rm cap}^{(2)} =\int_0^{\infty}\tau^2\rho(\tau)d\tau$ and $T^{(2)}$ is the second moment of the unconditional FPT density. Hence,
\begin{eqnarray}
FF_{k} &=&1-\overline{M}_k\bigg [\gamma(T_k+\tau_{\rm cap}) +\frac{\gamma}{2}\frac{T^{(2)}+2T\tau_{\rm cap}+\tau_{\rm cap}^{(2)}}{T+\tau_{\rm cap}}+O(\gamma^2)\bigg ] \nonumber .
\end{eqnarray}
It follows that
\begin{eqnarray}
\lim_{\gamma \rightarrow 0}FF_k =1- {\pi_k}\bigg [ \frac{T_k+\tau_{\rm cap}}{T+\tau_{\rm cap}} -\frac{1}{2}\frac{T^{(2)}+2T\tau_{\rm cap}+\tau_{\rm cap}^{(2)}}{(T+\tau_{\rm cap})^2}\bigg ].
\label{Fgam}
\end{eqnarray}
One other quantity that can be calculated without needing the individual fluxes $\widetilde{J}_{j}(\gamma)$ is the mean Fano factor per target:
 \begin{eqnarray}
 \label{meanFF}
  \langle{FF}\rangle:=\frac{1}{N}\sum_{k=1}^N FF_{k}=1+\frac{1}{N}\left [ \frac{1}{1-\widetilde{\rho}(\gamma)[1-\gamma \widetilde{Q}(\gamma)]}-1-N\langle{M}\rangle
\right ]\nonumber,
 \end{eqnarray}
 where
 \begin{equation}
 \label{Mbra}
 \langle{M}\rangle=\frac{1}{N}\sum_{k=1}^N\overline{M}_k=\frac{1}{\gamma N(T+\tau_{\rm cap})}.
 \end{equation}
 We see that $\langle FF\rangle$ depends on the survival probability and unconditional MFPT, both of which depend implicitly on $N$.

\setcounter{equation}{0}
\section{Diffusive search in a 3D domain with $N$ small interior targets}

Suppose that there are $N$ interior targets ${\mathcal U}_k$, $k=1,\ldots,N$, in a bounded domain ${\mathcal U}$ with a reflecting boundary $\partial{\mathcal U}$, rather than an absorbing exterior boundary as assumed in Fig. \ref{fig1}. The search domain is $\calU_c={\mathcal U}\backslash  \calU_a$ with boundary
$\partial \calU_c=\partial {\mathcal U}\cup \partial \calU_a$, where $\partial {\mathcal U}$ is reflecting and $\partial \calU_a$ absorbing. In general solving the FPT problem for $N$ targets is non-trivial even for simple geometric configurations. However, progress can be made if each target is taken to be sufficiently small, that is, $|\calU_k|=\epsilon^d|\calU|$ with $\epsilon \ll  1$ and $d=2,3$. We will also assume that the targets are well separated, in the sense that $|\x_i-\x_j|=O(1)$, $j\neq i$, and $\mbox{dist}(\x_j,\partial {\mathcal U})=O(1)$, where $\calU_j\rightarrow \x_j$ uniformly as $\epsilon \rightarrow 0$, $j=1,\ldots,N$. Under these conditions, one can use matched asymptotic expansions and Green's function methods \cite{Ward93,Coombs09,Cheviakov11,Chevalier11,Coombs15,Ward15,Lindsay16} to calculate the splitting probabilities $\pi_k$ and low-order moments of the conditional FPT densities. It is less straightforward to calculate the full FPT densities. However, as we show below, it is possible to calculate the Laplace transform of the survival probability. Once we have obtained these asymptotic expansions we can determine various statistical quantities, including the mean number of resources $\overline{M}_k$ according to equation (\ref{mean}), the small-$\gamma$ limit of the Fano factor $FF_k$ given by equation (\ref{Fgam}), and the mean Fano factor $\langle FF\rangle$ of equation (\ref{meanFF}). For the sake of illustration, we focus on 3D diffusion, although analogous methods can be used in 2D; the major difference is that the 2D Green's function has a logarithmic singularity \cite{Ward93,Ward15,Lindsay16}. 

\subsection{Survival probability}

The survival probability $Q(\x,t)$ defined in equation (\ref{q00}) evolves according to the backward diffusion equation
\begin{equation}
\frac{\partial Q(\x,t)}{\partial t} =D\nabla^2Q(\x,t) ,\ \x \in \calU,
\end{equation}
with  a reflecting boundary condition on the exterior of the domain
\begin{equation}
\partial_nQ(\x,t)=0,\quad \x\in \partial {\mathcal U},
\end{equation}
and absorbing boundary conditions on the target boundaries:
\begin{equation}
Q(\x,t)=0\quad \x \in \partial \calU_a =\bigcup_{j=1}^N\partial \calU_j.
\end{equation}
(For notational convenience, we drop the subscript on the initial position $\x_0$.)
The initial condition is $Q(\x,0)=1$. Laplace transforming the diffusion equation gives
\begin{equation}
D\nabla^2\widetilde{Q}(\x,s)-s \widetilde{Q}(\x,s)=-1,\ \x \in \calU,
\end{equation}
with the same boundary conditions. Following along the lines of Refs. \cite{Coombs09,Cheviakov11,Chevalier11,Coombs15}, we solve the boundary value problem for $\widetilde{Q}(\x,s)$ by constructing an inner or local solution
valid in an $O(\epsilon)$ neighborhood of each target, and then matching to an outer or global solution that is valid away from each neighborhood. One caveat is that the Laplace variable $s \ll 1/\epsilon^2$, otherwise the perturbation expansion in $\epsilon$ breaks down. 
In the outer region, which is outside an $O(\epsilon)$ neighborhood of each trap, $\widetilde{Q}(\x,s)$ is expanded as
\[\widetilde{Q}=\frac{1}{s}+\epsilon \widetilde{Q}_1+\epsilon^2 \widetilde{Q}_2+\ldots
\]
with
\begin{eqnarray}
\label{asym1}
D\nabla^2 \widetilde{Q}_n-s\widetilde{Q}_n&=0,\, \x\in \calU',\ \partial_n\widetilde{Q}_n=0,\, \x\in \partial \calU,
\end{eqnarray}
where $\calU'=\calU\backslash \{\x_1,\ldots,\x_N\}$, together with certain singularity conditions as $\x\rightarrow \x_j$, $j=1,\ldots,N$. The latter are determined by matching to the inner solution. In the inner region around the $j$-th target, we introduce the stretched coordinates ${\bf y}=\epsilon^{-1}(\x-\x_j)$ and set $q({\bf y},s) =\widetilde{Q}(\x_j+\epsilon \y,s)$. Expanding the inner solution as $q =q_0+\epsilon q_1+\ldots$,
we find that
\begin{eqnarray}
\label{inner}
\nabla_{\bf y}^2 q_n&=0,\ \y\in \R^d\backslash \calU_j \quad q_n(\y,s)=0,\ \y \in \partial \calU_j.
\end{eqnarray}
Finally, the matching condition is that the near-field behavior of the outer solution as $\x\rightarrow \x_j$ should agree with the far-field behavior of the inner solution as $|\y|\rightarrow \infty$, which is expressed as 
\[ \frac{1}{s}+\epsilon \widetilde{Q}_1+\epsilon^2 \widetilde{Q}_2+\ldots \sim q_0+\epsilon q_1+\ldots.
\]

First $q_0\sim s^{-1}$ so that we can set 
$ q_0(\y,s) =s^{-1}(1-w(\y))$, with $w(\y)$ satisfying the boundary value problem
\begin{eqnarray}
\label{w}
\nabla_{\bf y}^2 w(\y)&=&0,\  \y\in \R^d\backslash \calU_j ; \quad w(\y)=1,\ \y \in \partial \calU_j,\\
w(\y)&\rightarrow& 0\quad \mbox{as } |\y|\rightarrow \infty.\nonumber
\end{eqnarray}
This is a well-known problem in electrostatics and has the far-field behavior
\begin{equation}
w(\y)\sim \frac{C_j}{|\y|}+\frac{{\bf P}_j\cdot \y}{|\y|^3}+\ldots \mbox{as } |\y|\rightarrow \infty.
\end{equation}
where $C_j$ is the capacitance and ${\bf P}_j$ the dipole vector of an equivalent charged conductor with the shape $\calU_j$. (Here $C_j$ has units of length. In the case of a sphere of radius $l$ the capacitance is $C_j=l$.) It now follows that $\widetilde{Q}_1$ satisfies equation (\ref{asym1}) together with the singularity condition
\[\widetilde{Q}_1(\x,s)\sim -\frac{1}{s}\ \frac{C_j}{|\x-\x_j|} \quad \mbox{as } \x\rightarrow \x_j.\]
In other words, $\widetilde{Q}_1(\x,s)$ satisfies the inhomogeneous equation
\begin{eqnarray}
 D\nabla^2 \widetilde{Q}_1-s\widetilde{Q}_1&=\frac{4\pi D}{s} \sum_{j=1}^N C_j\delta(\x-\x_j),\, \x\in \calU',\quad
 \partial_n\widetilde{Q}_1=0,\ \x \in \partial \calU.
 \label{asym2}
\end{eqnarray}
This can be solved in terms of the modified Helmholtz Green's function
\begin{subequations}
\begin{eqnarray}
&&\nabla^2 G(\x,\x';\lambda_s)-\lambda_s^2 G(\x,\x';\lambda_s)= -\delta(\x-\x'),\, \x\in \calU;\\ &&\partial_nG=0,\, \x \in \partial \calU,\ \int_{\calU}G(\x,\x';\lambda_s)d\x=\lambda_s^{-2}\\
&&G(\x,\x';\lambda_s)=\frac{1}{4\pi|\x-\x'|}+H(\x,\x';\lambda_s),
\end{eqnarray}
\end{subequations}
with $\lambda_s=\sqrt{s/D}$ and $H(\x,\x';\lambda_s)$ corresponding to the regular (non-singular and boundary-dependent) part of the Green's function. Given $G$, the solution can be written as
\begin{equation}
 \widetilde{Q}_1(\x,s)=-\frac{4\pi}{s} \sum_{k=1}^NC_kG(\x,\x_k;\lambda_s).
\end{equation}

Next we match $q_1$ with the near field behavior of $\widetilde{Q}_1(\x,s)$ around the $j$-th target, which takes the form
\begin{eqnarray*}
\widetilde{Q}_1(\x,s)&\sim -\frac{1}{s} \frac{C_j}{|\x-\x_j|}-\frac{4\pi}{s} C_jH(\x_j,\x_j;\lambda_s) -\frac{4\pi}{s} \sum_{k\neq j}^NC_kG(\x_j,\x_k;\lambda_s).
\end{eqnarray*}
It follows that the far-field behavior is $q_1\sim {\chi_j}/{s}$ with
\begin{eqnarray}
 \chi_j&=&- {4\pi}  C_jH(\x_j,\x_j;\lambda_s) - {4\pi}  \sum_{k\neq j}^NC_kG(\x_j,\x_k;\lambda_s)=-4\pi\sum_{k=1}^NC_k{\mathcal G}_{jk}(s),\nonumber \\
\end{eqnarray}
where
${\mathcal G}_{ij}(s)=G(\x_i,\x_j;\lambda_s)$ for $i\neq j$, and ${\mathcal G}_{ii}(s)=H(\x_i,\x_i;\lambda_s)$.
The solution of equation (\ref{inner}) for $n=1$ is thus
\begin{equation}
q_1(\y,s)=\frac{\chi_j}{s}(1-w(\y)),\ {\bf y}=\epsilon^{-1}(\x-\x_j)
\end{equation}
with $w(\y)$ given by equation (\ref{w}). Hence, $\widetilde{Q}_2$ satisfies equation (\ref{asym1}) supplemented by the singularity condition
\[\widetilde{Q}_2(\x,s)\sim -\frac{1}{s}\ \frac{\chi_jC_j}{|\x-\x_j|} \quad \mbox{as } \x\rightarrow \x_j.\]
Following along identical lines to the derivation of $\widetilde{Q}_2(\x,s)$, we obtain the result
\begin{equation}
 \widetilde{Q}_2(\x,s)=-\frac{4\pi}{s} \sum_{k=1}^N\chi_kC_kG(\x,\x_k;\lambda_s).
\end{equation}
In conclusion, the outer solution takes the form
\begin{eqnarray}
\widetilde{Q}(\x,s)=\frac{1}{s}\bigg[1- {4\pi} \epsilon  \sum_{k=1}^NC_k(1+\epsilon \chi_k)G(\x,\x_k;\lambda_s)+o(\epsilon^2)\bigg ].
\label{QQQ0}
\end{eqnarray}

\subsection{Small-$s$ expansion and unconditional MFPT}

Taking the limit $s\rightarrow 0$ is non-trivial since there exist two small parameters, namely $s$ and $\epsilon$. First note that the Green's function has an expansion of the form
\begin{equation}
\label{GGG}
G(\x,\x';\sqrt{s/D})=\frac{D}{s|\calU|}+G(\x,\x')+O(s),
\end{equation}
where $G(\x,\x')$ is the Neumann Green's function for the diffusion equation:
\begin{subequations}
\label{G0}
\begin{eqnarray}
\nabla^2 G(\x;\x')=\frac{1}{|\calU|} -\delta(\x-\x'),\, \x\in \calU;\ \partial_nG=0,\, \x \in \partial \calU\\
G(\x,\x')=\frac{1}{4\pi|\x-\x'|}+H(\x,\x'),\ \int_{\calU}G(\x,\x')d\x=0,
\end{eqnarray}
\end{subequations}
with $H(\x,\x')$ corresponding to the regular part of the Green's function. Substituting into equation (\ref{QQQ0}) gives
\begin{eqnarray}
 \widetilde{Q}(\x,s)&=&\frac{1}{s}\bigg [1- {4\pi} \epsilon  \sum_{k=1}^NC_k\bigg (1+\epsilon \overline{\chi}_k-\Lambda \frac{\epsilon}{s}+\ldots \bigg ) \nonumber \\
 &&\quad \times \bigg (G(\x,\x_k)+\frac{D}{s|\calU|}+\ldots \bigg )+\ldots \bigg],\nonumber \\ 
\label{wow}
\end{eqnarray}
where
\begin{equation}
\label{Lame}
\Lambda=\frac{4\pi N\overline{C}D}{|\calU|},\quad \overline{C}=\frac{1}{N}\sum_{j=1}^N C_j,\quad 
\overline{\chi}_j=-4\pi\sum_{k=1}^NC_k\overline{\mathcal G}_{jk},
\end{equation}
with
$\overline{\mathcal G}_{ij}(s)=G(\x_i,\x_j)$ for $i\neq j$, and $\overline{\mathcal G}_{ii}(s)=H(\x_i,\x_i)$. Note that $\Lambda$ has units of inverse time.
Rearranging terms in equation (\ref{wow}), we can express $\widetilde{Q}$ as 
\begin{eqnarray}
\label{asymQ}
 \widetilde{Q}(\x,s)&=&\frac{1}{s}\left [1-\Lambda \left (\frac{\epsilon}{s}\right )+\Lambda^2\left (\frac{\epsilon}{s}\right )^2+\ldots \right ] \\
 &&\ - \left [ 4\pi \sum_{k=1}^NC_kG(\x,\x_k)\left (\frac{\epsilon}{s}\right ) +\frac{4\pi D}{\calU} \sum_{k=1}^NC_k\overline{\chi}_k\left (\frac{\epsilon}{s}\right )^2+\ldots\right ]+O(s^2).\nonumber
\end{eqnarray}
If we now take the limits $\epsilon,s\rightarrow 0$ with $\epsilon/s =1/\Lambda$ fixed, then $s\widetilde{Q}(\x,s)\rightarrow 1$, which implies that $Q(\x,t)\rightarrow 1$ as $t\rightarrow \infty$. Such a limit is consistent with the fact that the targets vanish in the zero-$\epsilon$ limit so the particle survives with probability one. It also suggests that setting $s=\Lambda \epsilon$ generates the correct $\epsilon$ expansion of $\widetilde{Q}(\x,0)$. Hence,
setting $s=\Lambda \epsilon$ in equation (\ref{asymQ}) yields the unconditional MFPT:
\begin{eqnarray}
\label{Tasym}
T \sim \frac{|\calU|}{4\pi \epsilon N D\overline{C}} \left [1-4\pi \epsilon \sum_{j=1}^NC_jG(\x_0,\x_j)+\frac{4\pi\epsilon}{N\overline{C}}  \sum_{i,j=1}^NC_i{\mathcal G}_{ij}C_j\right ].\nonumber
\end{eqnarray}
This reproduces the result obtained by directly solving the corresponding boundary value problem for $T$ \cite{Cheviakov11}.

\subsection{Splitting probabilities}

The splitting probabilities and conditional MFPTs can be analyzed in a similar fashion by writing down the appropriate boundary value problem and then matching inner and outer solutions \cite{Cheviakov11,Coombs15}. Here we simply summarize the result for the splitting probability $\pi_k(\x)$, which satisfies the boundary value problem
\begin{subequations}
\begin{eqnarray}
\nabla^2 \pi_k(\x)=0,\quad \x\in \calU; \quad \partial_n\pi_k(\x)=0,\quad \x \in \partial \calU,
\end{eqnarray}
with
\begin{equation}
\pi_k(\x)=1,\quad \x\in \partial \calU_k; \quad \pi_k(\x)=0 ,\quad \x \in \bigcup_{j\neq k} \partial \calU_j.
\end{equation}
\end{subequations}
The asymptotic expansion of the outer solution takes the form \cite{Cheviakov11}
\begin{eqnarray}
 \pi_k(\x)\sim \frac{C_k}{N\overline{C}}+4\pi \epsilon C_k\left [G(\x,\x_k)-\frac{1}{N\overline{C}}\sum_{j=1}^NC_jG(\x,\x_j)\right ] +\epsilon \chi_k+o(\epsilon),
\label{asympi3}
\end{eqnarray}
where
\begin{subequations}
\begin{equation}
\chi_k=-\frac{4\pi C_k}{N\overline{C}}\left [\sum_{j=1}^N\overline{\mathcal G}_{kj}C_j-\frac{1}{N\overline{C}}\sum_{i=1}^NC_i\overline{\mathcal G}_{ij} C_j\right ].
\end{equation}
\end{subequations}
 Note that $\sum_{k=1}^N\chi_k=0$. 

\subsection{Mean and Fano factor of resource distribution}

The above asymptotic results can now be used to determine various statistical quantities characterizing the distribution of resources. First, substituting equations (\ref{Tasym}) and (\ref{asympi3}) into equation (\ref{mean}) shows that the mean number of resources in the $k$-th target has the $\epsilon$ expansion
\begin{eqnarray}
\label{meank}
\overline{M}_k=\frac{\pi_k}{\gamma(T+\tau_{\rm cap})}=\frac{4\pi \epsilon  D C_k}{\gamma|\calU|(1+\epsilon\Lambda \tau_{\rm cap})}+o(\epsilon),
\end{eqnarray} 
with $\Lambda$ defined in equation (\ref{Lame}).
Hence, in the case of small targets, resources are more favorably delivered to targets with a larger shape capacitance $C_k$. Moreover, the leading-order contribution to the mean $\overline{M}_k$ is independent of the number of competing targets. Note that higher-order contributions depend on the geometry of the search domain and the positions of the other targets via the Green's function terms in equation (\ref{Tasym}).

Second, substituting the leading-order terms in the asymptotic expansions for $\pi_k$, $T$ and $T^{(2)}$ \cite{Coombs15} into equation (\ref{Fgam}) yields the small-$\gamma$ limit of the Fano factor $FF_k$:
\begin{eqnarray}
\lim_{\gamma \rightarrow 0}FF_k &=& 1-\frac{C_k}{N\overline{C}}\bigg [1 -\frac{1+\epsilon \Lambda \tau_{\rm cap}+\epsilon^2\Lambda^2\tau_{\rm cap}^{(2)}}{{(1+\epsilon\Lambda\tau_{\rm cap}})^2}\bigg ]+o(\epsilon)\nonumber \\
&=& 1-  \frac{4\pi \epsilon D\tau_{\rm cap}C_k}{|\calU|} +o(\epsilon).
\end{eqnarray}
Finally, substituting equation (\ref{QQQ0}) into the expression (\ref{meanFF}) for the target-averaged Fano factor gives
 \begin{eqnarray}
 \langle{FF}\rangle
 = 1+\frac{{4\pi} \epsilon \widetilde{\rho}(\gamma)}{N}   \sum_{k=1}^NC_k G(\x,\x_k;\lambda_{\gamma}) -\frac{4\pi \epsilon \overline{C}D}{\gamma |\calU|}+o(\epsilon).
 \label{oh}
\end{eqnarray}
Hence, the Fano factors have $O(\epsilon)$ deviations from unity that depend on the distribution of targets within the domain $|\calU|$.

\setcounter{equation}{0}

\section{Stochastic resetting before capture} 

\begin{figure}[h!]
\raggedleft
  \includegraphics[width=7cm]{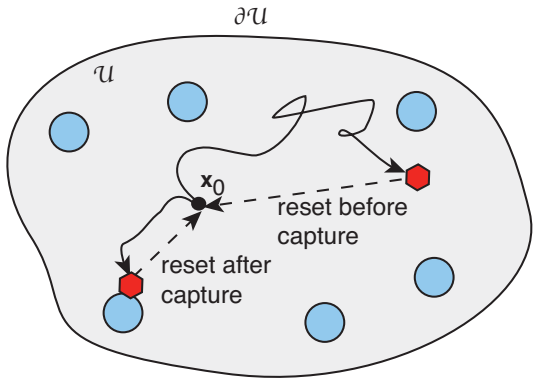}
  \caption{Random search in a bounded domain $\calU$ with totally absorbing targets and two types of resetting event: ``reset after capture'' and ``reset before capture.'' }
  \label{fig8}
\end{figure}

Now suppose that prior to being absorbed by one of the targets, the particle can reset to the initial position $\x_0$ at a random sequence of times generated by an exponential probability density $\psi(\tau)=r\e^{-r\tau}$, where $r$ is the resetting rate. The probability that no resetting has occurred up to time $\tau$ is then $\Psi(\tau)=1-\int_0^{\tau}\psi(s)ds=\e^{-r\tau}$. Note that, in contrast to resetting after capture, the sequence of resetting times is generated independently of the ongoing search process. Following a resetting event the particle immediately returns to $\x_0$ and restarts the current search. For simplicity, we will assume that the major source of delay is due to the loading/unloading of cargo so that finite return times and refractory periods associated with resetting before capture are ignored. The two types of resetting event are illustrated in Fig. \ref{fig8}.  

A number of authors have recently calculated the splitting probabilities and conditional MFPTs of a single search-and-capture event in the presence of stochastic resetting and two or more targets \cite{Belan18,Chechkin18,Pal19,Bressloff20m}, extending previous work on single targets \cite{Reuveni16,Pal17,Pal20,Bodrova20}. (Several of these studies also allow for non-exponential resetting and finite return times, which we do not consider in this paper.) The basic idea is to exploit the fact that once the particle has returned to $\x_0$ it has lost all memory of previous search phases, which means that one can condition on whether or not the particle resets at least once, even though a reset event occurs at random times. Renewal theory can then be used to express statistical quantities with resetting in terms of statistical quantities without resetting. In order to distinguish between the two cases, we will add a subscript $r$ to the splitting probabilities etc. of the former.

\subsection{Splitting probabilities and MFPTs}

\begin{figure}[t!]
\raggedleft
  \includegraphics[width=10cm]{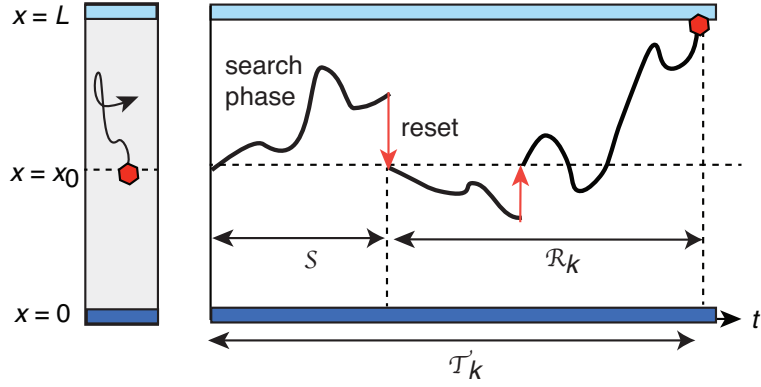}
  \caption{Single search-and-capture event with stochastic resetting before capture for a pair of exterior targets located at the ends of the finite domain $x\in [0,L]$. Sample trajectory of the particle finding the second target after two resetting events. Following each resetting event, the particle immediately returns to the point $\x_0$, after which it re-enters the search phase. Also shown is the decomposition of the conditional FPT to the $k$-th target, $k=1,2$, ${\mathcal T}_k={\mathcal S}+{\mathcal R}_k$. See text for details.}
  \label{fig9}
\end{figure}

Let ${\mathcal I}(t)$ denote the number of resettings in the interval $(0,t)$. Consider the following set of first passage times, see Fig. \ref{fig9}:
\begin{eqnarray}
{\mathcal T}_k&=\inf\{t\geq 0; \X(t)\in \partial \calU_k,  {\mathcal I}(t)\geq 0\},\nonumber \\
\label{defT}
{\mathcal S}&=\inf\{t\geq 0; \X(t)=\x_0,\ {\mathcal I}(t) =1 \},\\
{\mathcal R}_k &=\inf\{t\geq 0; \X(t+{\mathcal S})\in \partial \calU_k,{\mathcal I}(t+{\mathcal S})\geq 1 \}.\nonumber
\end{eqnarray}
Here ${\mathcal T}_k$ is the FPT for finding the $k$-th target irrespective of the number of resettings, ${\mathcal S}$ is the FPT for the first resetting and return to $\x_0$ without being captured by any target, and ${\mathcal R}_k$ is the FPT for finding the $k$-th target given that at least one resetting has occurred. Next we define the sets $\Omega_k = \{ \mathcal{T}_k< \infty \}$ and $\Gamma_k = \{ \mathcal{S} < \mathcal{T}_k< \infty\}\subset \Omega_k$, 
where $\Omega_k$ is the set of all events for which the particle is eventually absorbed by the $k$-th target without being absorbed by any other target, and $\Gamma_k$ is the subset of events in $\Omega_k$ for which the particle resets at least once. It immediately follows that $\Omega_k \backslash \Gamma_k = \{ \mathcal{T} _k< {\mathcal S} = \infty \}$, 
where $\Omega_k\backslash \Gamma_k$ is the set of all events for which the particle is captured by the $k$-th target without any resetting. 

Given the above definitions, the splitting probability $\pi_{r,k}$ can be decomposed as 
\begin{equation} 
\label{rho1}
\pi_{r,k}(\x_0):= \mathbb{P}[\Omega_k] =  \mathbb{P}[\Omega_k\backslash\Gamma_k]+ \mathbb{P}[\Gamma_k].
\end{equation}
Suppose that the splitting probability $\pi_{r,k}$ is decomposed according to equation (\ref{rho1}):
\begin{equation}
\pi_{r,k}(\x_0):= \mathbb{P}[\Omega_k] =  \mathbb{P}[\Omega_k\backslash\Gamma_k]+ \mathbb{P}[\Gamma_k].
\end{equation}
The probability that the particle is captured by the $k$-th target in the interval $[\tau,\tau+d\tau]$ without any returns to $\x_0$ is $\Psi(\tau)J_k(\x_0,\tau)d\tau$ with $J_k(\x_0,\tau)$ given by equation (\ref{Jk}). Hence,
\begin{eqnarray}
\P[\Omega_k\backslash\Gamma_k]&= \int_0^{\infty}\Psi(\tau) J_k(\x_0,\tau)d\tau=-\int_0^{\infty}\e^{-r\tau}\frac{d\Pi_k(\x_0,\tau)}{d\tau} d\tau \nonumber \\
&=\pi_{k}(\x_0)-r\widetilde{\Pi}_{k}(\x_0,r),
\label{PS0}
\end{eqnarray}
after integrating by parts.
Next, from the definitions of the first passage times, we have
\begin{equation}
\label{bobo}
\mathbb{P}[\Gamma_k]=\P[{\mathcal S}<\infty]\P[{\mathcal R}_k<\infty],
\end{equation}
and memoryless return to $\x_0$ implies that $\P[{\mathcal R}_k<\infty]=\pi_{r,k}(\x_0)$. In addition
\begin{eqnarray}
  \P[{\mathcal S}<\infty]&=r\int_0^{\infty}\e^{-r\tau}Q(\x_0,\tau)d\tau=r\widetilde{Q}(\x_0,r)=r\sum_{k=1}^N \widetilde{\Pi}_{k}(\x_0,r) .
\label{PS}
\end{eqnarray}
We have used equation (\ref{q00}) and the fact that the probability of resetting in the time interval $[\tau,\tau + d\tau]$ is equal to the product of the reset probability $\psi(\tau)d\tau$ and the survival probability $Q(\x_0,t)$. Hence, equation (\ref{bobo}) becomes
\begin{equation}
\label{PG}
\P[\Gamma_k]=\pi_{r,k}(\x_0)r\widetilde{Q}(\x_0,r) .
\end{equation}
Finally, combining equations (\ref{PS0}) and (\ref{PG}) and rearranging gives
\begin{eqnarray}
\pi_{r,k}(\x_0)&=&\frac{\pi_{k}(\x_0)-r\widetilde{\Pi}_{k}(\x_0,r)}{ 
1-r\widetilde{Q}(\x_0,r)}=\frac{\widetilde{J}_k(\x_0,r)}{ 
1-r\widetilde{Q}(\x_0,r)}.
\label{Piee}
\end{eqnarray}

A similar analysis can be carried out for the decomposition of the conditional FPT densities :
\begin{eqnarray}
\label{M1}
\pi_{r,k}(\x_0)\widetilde{f}_{r,k}(\x_0,s)&=	\E[\e^{-s{\mathcal T}_k}1_{\Omega_k}] = \mathbb{E}[\e^{-s{\mathcal T}_k}1_{\Omega_k \backslash \Gamma_k}]+ \mathbb{E}[\e^{-s{\mathcal T}_k}1_{\Gamma_k}].\nonumber
\end{eqnarray}
The first expectation can be evaluated by noting that it is the FPT density for capture by the $k$-th target without any resetting, and the probability density for such an event is $\Psi(\tau)J_k(\x_0,\tau)d\tau$: 
\begin{eqnarray}
 \label{M2}
  \mathbb{E}[\e^{-s{\mathcal T}_k}1_{\Omega_k\backslash \Gamma_k}] &= &\int_0^{\infty}\e^{-(s+r)\tau}J_k(\x_0,\tau)d\tau=-\int_0^{\infty}\e^{-(s+r)\tau}\frac{d\Pi_k(\x_0,\tau)}{d\tau} d\tau \nonumber  \\
 &=&\pi_{k}(\x_0)-(r+s)\widetilde{\Pi}_{k}(\x_0,r+s).\nonumber
 \end{eqnarray}
The second expectation can be written as
\begin{eqnarray}
 \mathbb{E}[\e^{-s{\mathcal T}_k}1_{\Gamma_k}] 
	&=&\mathbb{E}[\e^{-s[{\mathcal S}+{\mathcal R}_k]}1_{\Gamma_k}]\nonumber \\
	&=&\left (\int_0^{\infty}r\e^{-(r+s)\tau_1}Q(\x_0,\tau_1)d\tau_1 \right )\pi_{r,k}(\x_0)\widetilde{f}_{r,k}(\x_0,s).\nonumber \\
	\label{M3}
	\end{eqnarray} 
We have used the fact that the probability that the first return is initiated in the interval $[\tau_1,\tau_1+d\tau_1]$, given that $\X(\tau_1)=\x$ and the particle has not been captured by a target, is $\psi(\tau)p_0(\x,\tau|\x_0)d\tau_1$.  The particle then immediately returns to $\x_0$ and restarts the search. The remaining time to find the $k$-th target has the same conditional first passage time density as ${\mathcal T}_k$.
Combining equations (\ref{M1})--(\ref{M3}) and rearranging yields the result
\begin{eqnarray}
\label{Tcond1}
\pi_{r,k}(\x_0)\widetilde{f}_{r,k}(\x_0,s)=\frac{\pi_{k}(\x_0)-(r+s)\widetilde{\Pi}_{k}(\x_0,r+s)}{1-r\widetilde{Q}(\x_0,r+s)},
\end{eqnarray}

The Laplace transform of the FPT density is the moment generator of the conditional FPT ${\mathcal T}_k$:
\begin{equation}
T_{r,k}^{(n)}=\E[{\mathcal T}_k^n1_{\Omega_k}]=\left .\left (-\frac{d}{ds}\right )^n\E[\e^{-s{\mathcal T}_k}1_{\Omega_k}]\right |_{s=0}.
\end{equation}
For example, the conditional MFPT $T_{r,k}=T_{r,k}^{(1)}$ is
\begin{eqnarray}
\label{Tcond2}
 \pi_{r,k}(\x_0)T_{r,k}(\x_0)  &=\frac{\widetilde{\Pi}_{k}(\x_0,r) +r\widetilde{\Pi}_{k}'(\x_0,r)}
{1-
r\widetilde{Q}(\x_0,r)}\  -\pi_{r,k}(\x_0)\left [\frac{r\sum_k\widetilde{\Pi}_k'(\x_0,r)}{ 
1-r\widetilde{Q}(\x_0,r)}\right ].  
\end{eqnarray}
where $'$ denotes differentiation with respect to $r$.
Finally, summing equation (\ref{Tcond2}) with respect to $k$ implies that the unconditional MFPT is simply
\begin{eqnarray}
\label{TQ}
T_r(\x_0)&:=\sum_{k=1}^N\pi_{r,k}(\x_0)T_{r,k}(\x_0)=\frac{\widetilde{Q}(\x_0,r)}
{1-
r\widetilde{Q}(\x_0,r)}.
\end{eqnarray}

\subsection{Analysis of mean and variance}

We can now use the results of sections 2 to determine the mean and Fano factor of the resource distribution in the presence of resetting:
\begin{equation} \label{mean2}
\overline{M}_{r,k}  = \frac{\pi_{r,k}}{\gamma   (T_{r} +\tau_{\rm cap} )},
\end{equation}
and
\begin{eqnarray}
\label{varr}
 \mbox{Var}[M_{r,k}]=\overline{M}_{r,k} \left [ \frac{\pi_{r,k}\widetilde{\rho}(\gamma)\widetilde{f}_{r,k}(\gamma)}{1-\widetilde{\rho}(\gamma)\sum_{j=1}^N\pi_{r,j} \widetilde{ f}_{r,j}(\gamma)}+1-\overline{M}_{r,k} 
\right ],
\end{eqnarray}
with $\pi_{r,k}$ and $T_{r}$ given by equations (\ref{Piee}) and (\ref{TQ}), respectively. It turns out that we can express the variance of the target resource distribution in a particularly useful form, analogous to corresponding results for a single target \cite{Reuveni16,Bressloff20q}. 
That is, rearranging equation (\ref{TQ}) gives
\[\widetilde{Q}(\x_0,r)=\frac{T_r(\x_0)}{1+r T_r(\x_0)}.\]
which on substituting into equation (\ref{Piee}), leads to the result
\begin{eqnarray}
\frac{\pi_{r,k}(\x_0)}{\pi_k(\x_0)}\frac{1}{1+rT_r(\x_0)} =1-r\frac{\widetilde{\Pi}_{k}(\x_0,r)}{\pi_k(\x_0)}.
\end{eqnarray}
It follows that equation (\ref{Tcond1}) can be rewritten as
\begin{eqnarray}
\label{ffLT2}
 \pi_{r,k}(\x_0)\widetilde{f}_{r,k}(\x_0,s)&=&\pi_k(\x_0)\left [ \frac{\pi_{r+s,k}(\x_0)}{\pi_k(\x_0)} \frac{1}{1+(r+s)T_{r+s}(\x_0)}\right ]\nonumber
  \\
  &&\quad \times \left [1-\frac{r T_{r+s}(\x_0)}{1+(r+s) T_{r+s}(\x_0)}\right ]^{-1} \nonumber \\
   &&= \frac{\pi_{r+s,k}(\x_0)}{1+s T_{r+s}(\x_0)}.
\end{eqnarray}
Finally, substituting equation (\ref{ffLT2}) into equation (\ref{varr}), we have
(after dropping the explicit dependence on $\x_0$)
\begin{eqnarray}
\label{var2}
 \mbox{Var}[M_{r,k}]
= \overline{M}_{r,k} \left [\frac{  {\pi}_{r+\gamma,k}\widetilde{\rho}(\gamma)}{\gamma T_{r+\gamma}+1-\widetilde{\rho}(\gamma)}-\overline{M}_{r,k}+1\right ] .
\end{eqnarray}
Note that in the absence of loading/unloading delays, $\widetilde{\rho}(\gamma)=1$ and $\tau_{\rm cap}=0$, we obtain the simple expression
\begin{eqnarray}
\label{var2a}
& \mbox{Var}[M_{r,k}]
 =\overline{M}_{r,k} \left [\overline{M}_{r+\gamma,k}-\overline{M}_{r,k} +1 \right ] .
\end{eqnarray}
That is, the variance of resource accumulation with exponential resetting and no delays is determined completely in terms of the means $ \overline{M}_{r,k} $ and $ \overline{M}_{r+\gamma,k}$. An analogous result was previously obtained for a single target \cite{Bressloff20q}. 

Again it is more convenient to work with the Fano factor with resetting, which is given by
\begin{equation}
\label{FFr}
FF_{r,k}=\frac{\mbox{Var}[M_{r,k}]}{\overline{M}_{r,k}}= \frac{  {\pi}_{r+\gamma,k}\widetilde{\rho}(\gamma)}{\gamma T_{r+\gamma}+1-\widetilde{\rho}(\gamma)}-\overline{M}_{r,k}+1  .
\end{equation}
Summing both sides with respect to $k$ then yields the mean Fano factor per target,
\begin{equation}
\label{FFmeanr}
\langle FF_{r}\rangle =\frac{1}{N}\sum_{k=1}^MFF_{r,k}=\frac{1}{N} \frac{ \widetilde{\rho}(\gamma)}{\gamma T_{r+\gamma}+1-\widetilde{\rho}(\gamma)}-\langle {M}_{r}\rangle+1  ,
\end{equation}
where
\begin{equation}
\label{Mrbra}
 \langle{M}_r\rangle=\frac{1}{N}\sum_{k=1}^N\overline{M}_{r,k}=\frac{1}{\gamma N(T_r+\tau_{\rm cap})}.
 \end{equation}
Suppose that we fix the time scale by setting $\gamma=1$, and vary the resetting rate $r$. Equations (\ref{mean2}) and (\ref{var2}) imply that 
the mean and variance both vanish in the large-$r$ limit; in the case of fast resetting, the particle rarely has the chance to deliver resources and so degradation dominates and $T_r\rightarrow \infty$. Moreover, 
\begin{equation}
\label{conr}
\lim_{r \rightarrow \infty} FF_{r,k}=1,\quad \lim_{r \rightarrow 0} FF_{r,k}<\infty.
\end{equation}

\subsection{Diffusive search in 3D}

We now consider the effects of resetting on diffusive search in a 3D domain with small targets, see section 3. In order to calculate the corresponding quantities under resetting, we need to determine the Laplace transformed fluxes $\widetilde{J}_k(\x_0,r)$, see equations (\ref{Piee}) and (\ref{Tcond2}), which is not straightforward. One approach is to carry out a perturbation expansion of $\widetilde{J}_k(\x_0,r)$, which provides information regarding the distribution of resources in the small-$r$ regime \cite{Bressloff20m}.
Here we consider the simpler problem of calculating the mean number of resources and the Fano factor averaged over the $N$ targets. 

First, substituting equation (\ref{QQQ0}) into equation (\ref{TQ}) gives
\begin{eqnarray}
\label{TQ2}
T_r
&=&\frac{1}{4\pi \epsilon r}\frac{1-4\pi \epsilon \sum_{k=1}^NC_kG(\x_0,\x_k;\lambda_r)+o(\epsilon)}{\sum_{k=1}^NC_k(1+\epsilon \chi_k)G(\x_0,\x_k;\lambda_r)+o(\epsilon)} \nonumber\\
&=&\frac{1}{4\pi \epsilon r\sum_{k=1}^NC_kG(\x_0,\x_k;\lambda_r)} +O(1).
\end{eqnarray}
Note that $T_r\rightarrow \infty$ as $\epsilon\rightarrow \infty$, as expected in the limit of vanishing targets. Second, substituting for $T_r$ into equation (\ref{Mrbra}) with $\tau_{\rm cap}=0$ gives the leading order expression
\begin{eqnarray} \label{means}
\langle M_r\rangle   =\frac{1}{N\gamma T_{r} }=\frac{4\pi \epsilon r\sum_{k=1}^NC_kG(\x_0,\x_k;\lambda_r)} {N\gamma }+o(\epsilon).
\end{eqnarray}
Note that for small $r$, we can use equation (\ref{GGG}) so that
\begin{eqnarray}
\langle M_r\rangle   \approx\frac{4\pi \epsilon \overline{C}D}{\gamma |\calU|}+\frac{4\pi \epsilon r\sum_{k=1}^NC_kG(\x_0,\x_k)} {N\gamma }+o(r,\epsilon).
\end{eqnarray}
This implies that the introduction of slow resetting increases the mean number of resources per target if and only if $\sum_{k=1}^NC_kG(\x_0,\x_k;0)>0$. This agrees with a previous analysis based on a small-$r$ expansion \cite{Bressloff20m}. Moreover, we recover the $r=0$ result given by summing equation (\ref{meank}) over $k$. 
Finally, turning to equation (\ref{FFmeanr}) for the mean Fano factor per target, we have
 \begin{eqnarray}
 \langle{FF}_r\rangle 
& =&1-\frac{4\pi \epsilon r\sum_{k=1}^NC_kG(\x_0,\x_k;\lambda_r)} {N\gamma } \\
 &&\quad +\frac{4\pi \epsilon \widetilde{\rho}(\gamma)(r+\gamma)\sum_{k=1}^NC_kG(\x_0,\x_k;\lambda_{r+\gamma})}{\gamma N}+o(\epsilon).\nonumber 
 \end{eqnarray}
 Thus deviations of the target-averaged Fano factor from unity are $O(\epsilon)$ and depend on the distribution of targets within the domain $\calU$ via the Green's function terms. In the limit $r\rightarrow 0$ we recover equation (\ref{oh}).

\section{Discussion} 

In this paper we investigated resource accumulation in a population of targets under multiple rounds of diffusive search-and-capture. The boundary of each target within the search domain was taken to be totally absorbing. However, following target capture, we assumed that the particle unloads a resource packet and then returns to its initial position, where it is reloaded with cargo and initiates a new search process (resetting after capture). We then used asymptotic analysis to investigate the distribution of resources in a population of small targets in a 3D domain. In particular, we expressed various statistical quantities as asymptotic perturbation expansions in the target size $\epsilon$. We thus showed that the mean number of resources in a target depends on its shape capacitance, while the corresponding Fano factor has $O(\epsilon)$ deviations from unity that depend on the spatial locations of the targets. Finally, we extended our results to include the effects of stochastic resetting before capture, under the additional assumption that the primary source of delays arises from the resource loading/unloading times.

It is important to emphasize that the theoretical framework developed in this paper can be applied to more general search-and-capture processes such as velocity-jump processes and advection-diffusion equations; one can also include the effects of finite return times and non-exponential resetting. Velocity jump processes are often used to model motor-driven active transport processes in cells, where the particle randomly switches between left-moving and right-moving velocity states. Moreover, in the limit of fast switching, a quasi-steady-state approximation can be used to reduce the transport equation to an advection-diffusion equation \cite{Newby10}. 

Finally, note that in this paper we focused on a single searcher, whereas a more common scenario is to have many parallel searchers. However, our results carry over to this case provided that the searchers are independent. That is, suppose that there are ${\mathcal N}$ independent, identical searchers. Statistical independence implies that both the steady-state mean and variance scale of resources within a target scale as ${\mathcal N}$. Hence, the Fano factor $FF_k$ is independent of ${\mathcal N}$, whereas the coefficient of variation scales as $CV_k\sim 1/\sqrt{\mathcal N}$. The latter indicates that the size of fluctuations decreases as the number of searchers increases, which is also a manifestation of the law-of-large numbers.


\bigskip





\begin{thebibliography}{10}

\bibitem{Bell91} Bell W J 1991 {\em Searching behaviour: the behavioural ecology of finding resources}. Chapman and
Hall, London


\bibitem{Bartumeus09} Bartumeus F and Catalan J 2009 Optimal search behaviour and classic foraging theory. {\em J. Phys. A: Math. Theor.} {\bf 4} 434002.

\bibitem{Viswanathan11} Viswanathan G M, da Luz M G E, Raposo E P and Stanley H E 2001 {\em The Physics of Foraging: An Introduction to Random Searches and Biological Encounters.} Cambridge University Press.


\bibitem{Berg81} Berg O G, Winter R B and von Hippel P H 1981 Diffusion-driven mechanisms of protein translocation on nucleic acids. I. Models and theory. {\em Biochemistry} {\bf 20} 6929.

\bibitem{Halford04}
Halford S E and Marko J F 2004 How do site-specific {DNA}-binding proteins find
  their targets?
\newblock {\em Nucl. Acid Res.} \textbf{32} 3040-3052.

\bibitem{Coppey04} Coppey M, Benichou O, Voituriez R and Moreau M 2004 Kinetics of target site localization of a protein on DNA: A stochastic approach. {\em Biophys. J.} {\bf 87} 1640.

\bibitem{Lange15} Lange M, Kochugaeva M and Kolomeisky A B 2015 Protein search for multiple targets on DNA. {\em J. Chem. Phys.} {\bf 143} 105102.

\bibitem{Loverdo08}
Loverdo C, Benichou O, Moreau M, Voituriez R 2008 Enhanced reaction
  kinetics in biological cells {\em Nat. Phys.} \textbf{4} 134-137
  
  \bibitem{Benichou10} Benichou O, Chevalier C, Klafte J, Meyer B and Voituriez R 2010 Geometry-controlled
kinetics. {\em Nat. Chem.} {\bf 2} 472-477.
  

 \bibitem{Bressloff13} Bressloff P C and Newby J M 2013 Stochastic models of intracellular transport. {\em Rev. Mod. Phys.} {\bf 85} 135-196.

 
\bibitem{Maeder14} Maeder C I, San-Miguel A, Wu E Y, Lu H and Shen K 2014 {\em Traffic} {\bf 15} 273-291

 
\bibitem{Bressloff15} Bressloff P C and Levien E 2015 Synaptic democracy and active intracellular transport in axons. {\em Phys. Rev. Lett.} {\bf 114} 168101  



\bibitem{Kornberg14} Kornberg T B and Roy S 2014 Cytonemes as specialized
signaling filopodia. {\em Development} {\bf 141} 729-736

\bibitem{Stanganello16} Stanganello E and Scholpp S 2016 Role of cytonemes
in Wnt transport {\em J. Cell Sci.} {\bf 129} 665-672


 \bibitem{Zhang19} 
 Zhang C and Scholpp S 2019 {Cytonemes in development.} {\em Curr. Opin. Gen. Dev.} {\bf 58} 25-30.

 
  \bibitem{Bressloff19} Bressloff P C and Kim H 2019 A search-and-capture model of cytoneme-mediated morphogen gradient formation. {\em Phys. Rev. E} {\bf 99} 052401.
  
  
 \bibitem{Bressloff20a} Bressloff P C 2020 Modeling active cellular transport as a directed search process with stochastic resetting and delays {\em J. Phys. A: Math. Theor.} {\bf 53} 355001.


 


\bibitem{Takacs62} Takacs L 1962 {\em Introduction to the theory of queues.} Oxford University Press, Oxford.

\bibitem{Liu90} Liu L, Kashyap B R K and Templeton J G C 1990 On the GIX/G/Infinity system. {\em J. Appl Prob.} {\bf 27} 671-683.

\bibitem{Redner} Redner S 2001 {\em A Guide to First-Passage Processes}. Cambridge University Press, Cambridge, UK.






\bibitem{Evans20} Evans M R, Majumdar S N, Schehr G 2020 Stochastic resetting and applications {\em J. Phys. A: Math. Theor.} {\bf 53} 193001.
 



\bibitem{Little61} Little J D C 1961 A Proof for the Queuing Formula: $L=\lambda W$. {\em Operations Research}. {\bf 9} 383-387.



\bibitem{Coombs09} Coombs D, Straube R and  Ward M 2009 Diffusion on a sphere with localized traps: Mean first passage time, eigenvalue asymptotics, and Fekete points. {\em SIAM J. Appl. Math.} {\bf 70} 302-332.

\bibitem{Cheviakov11} Cheviakov A F and Ward M J 2011 Optimizing the principal eigenvalue of the Laplacian in a sphere with interior traps. {\em Math.
Comp. Modeling} {\bf 53} 1394-1409.



\bibitem{Chevalier11} Chevalier C, Benichou O, Meyer B and Voituriez R 2011 First-passage quantities of Brownian motion in a bounded domain with multiple targets: a unified approach. {\em J. Phys. A} {\bf 44} 025002.

\bibitem{Coombs15} Delgado M I, Ward M and Coombs D 2015 Conditional mean first passage times to small traps in a 3-D domain with a sticky boundary: Applications to T cell searching behavior in lymph nodes. {\em Multiscale Model. Simul.} {\bf 13} 1224-1258.


\bibitem{Ward93} Ward M J, Henshaw W D and Keller J B 1993 Summing logarithmic expansions for singularly perturbed eigenvalue problems {\em SIAM J. Appl. Math} {\bf 53} 799-828 

\bibitem{Ward15} Kurella V, Tzou J C, Coombs D and Ward M J 2015
{Asymptotic analysis of first passage time problems
inspired by ecology.} {\em Bull Math Biol.} {\bf 77} 83-125.

\bibitem{Lindsay16} Lindsay A E, Spoonmore R T and Tzou J C 2016 Hybrid asymptotic-numerical approach for estimating first passage time densities of the two-dimensional narrow capture problem. {\em Phys. Rev. E} {\bf 94} 042418.



\bibitem{Belan18} Belan S 2018 Restart could optimize the probability of success in a Bernouilli trial. {\em Phys. Rev. Lett.} {\bf 120} 080601

\bibitem{Chechkin18} Chechkin A and Sokolov I M 2018 Random search with resetting: A unified renewal approach. {\em Phys. Rev. Lett.} {\bf 121} 050601.


\bibitem{Pal19} Pal A and Prasad V V 2019 First passage under stochastic resetting in an interval. {\em Phys. Rev. E} {\bf 99} 032123

\bibitem{Bressloff20m} Bressloff P C 2020 Search processes with stochastic resetting and multiple targets. {\em Phys. Rev. E} {\bf 102} 022115.


\bibitem{Reuveni16} Reuveni S 2016 Optimal stochastic restart renders fluctuations in first-passage times universal {\em Phys. Rev. Lett.} {\bf 116} 170601

  \bibitem{Pal17} Pal A and Reuveni S 2017 First passage under restart {\em Phys.
Rev. Lett.} {\bf 118}, 030603 


\bibitem{Pal20} Pal A,  Kusmierz L, Reuveni S. 2020. Home-range search provides advantage under high uncertainty. {\em arXiv:1906.06987} (2020).


\bibitem{Bodrova20} Bodrova A S, Sokolov I M 2020 Resetting processes with noninstantaneous return {\em Phys. Rev. E} {\bf 101} 052130


\bibitem{Bressloff20q} Bressloff P C 2020 Queueing theory of search processes with stochastic resetting. {\em Phys. Rev. E} Submitted.

\bibitem{Newby10} Newby J and Bressloff P C 2010 Quasi-steady state reduction of molecular-based models of directed intermittent search. {\em Bull. Math. Biol.} {\bf 72} 1840.



\end{thebibliography}
\end{document}